\documentclass[twocolumn,showpacs,preprintnumbers,amsmath,amssymb]{revtex4}
\usepackage{graphicx}

\begin{document}
\title{{\Large  Pairing and shell evolution in neutron rich nuclei}}

\author{M. Saha Sarkar
\thanks{e-mail: maitrayee.sahasarkar@saha.ac.in}}
\affiliation{Nuclear Physics Division, Saha Institute
of Nuclear Physics, Kolkata 700064, INDIA }

\author{S. Sarkar
\thanks{e-mail: ss@physics.becs.ac.in}}
\affiliation{Department   of   Physics,  Bengal  Engineering  and
Science University, Shibpur, Howrah - 711103, INDIA}

\date{\today}

\begin{abstract}

From  the  experimental data on odd-even staggering of masses, we
have shown that variation of pairing as  a  function  of  neutron
number  plays an important role in many distinctive features like
occurrence  of  new  shell  closures,  shell  erosion,   anomalous
reduction  of  the  energy  of  the  first 2$^+$ state and slower
increase  in  the  B($E2,  2^+_1  \rightarrow  0^+_1$)   in   the
neutron-rich  even-even  nuclei  in  different  mass regions. New
predictions have been made in a model independent way. Results of
theoretical calculations support the phenomenological findings.

\end{abstract}

\maketitle

The  energy  of  the  first 2$^+$ state ($E(2^+_1)$) of even-even
nucleus is a sensitive probe to  study  the  evolution  of  shell
structure. For spherical nuclei, this state is formed by breaking
a  pair  spending  the  pair  binding  energy, showing pronounced
maxima at the shell closures. Nuclei away from the shell closures
are  gradually  driven  away  from  spherical   symmetry,   where
"anomalously"  low  first  excited  collective  2$^+$  states are
observed.

Usually,  a  few distinctive features have been identified as the
signatures  of  changing  shell  structure  away  from  stability
\cite{sorlin}.  Three  new doubly magic Oxygen isotopes have been
observed.  For  neutron  rich  nuclei,   semi-magic   ones   like
$^{32}_{12}Mg_{20}$,  $^{30}_{10}Ne_{20}$  have  shown erosion of
N=20 shell closure  with  sudden  decrease  in  their  $E(2^+_1)$
values  and  an  increase  in  their  corresponding  B($E2, 2^+_1
\rightarrow 0^+_1$). However some recent measurements of  reduced
($E(2^+_1)$)  and  B(E2)  values  in highly neutron nuclei having
Z=4-18 have raised a serious discussion on hindered  E2  strength
unexpected   for  these  nuclei  phenomenologically  as  well  as
theoretically \cite{raman,ref,be,Ar}. In  heavier  nuclei,  after
observation  \cite{rad} of reduction of both $E(2^+_1)$ and B(E2)
in $^{136}Te$, Terasaki {\it et  al.}  \cite  {tera},  from  QRPA
calculations,  traced  the  origin of this anomalous behaviour in
$^{136}Te$ isotope to a reduced neutron pairing above the  N$=$82
magic   gap.  Around  this  period,  new  empirical  interactions
\cite{epja}  proposed  for  neutron  rich  isotopes   above   the
$^{132}Sn$  core  and  later those for neutron rich nuclei in the
$sd-fp$ shell \cite{signo:11, poves} also included  reduction  of
pairing matrix elements for better reproduction of data. So these
studies indicate an important role of pairing in the evolution of
the structure of exotic nuclei.

In  this  letter,  we  show  from  the neutron and proton pairing
energies estimated from the differences in  experimental  binding
energies  in different mass regions, that pairing plays a crucial
role in evolution  of  shell  structure  in  atomic  nuclei.  For
observed  shell erosions at N=8 in Be isotopes and at N=20 in Ne,
Mg isotopes, relative enhancement of proton  pairing  along  with
weakening  of neutron pairing have been found to be important. We
have found a clear correlation among the occurrence of  new  shell
closures,  the  reduction of $E(2^+_1)$ and hindered B(E2) in the
neutron-rich even-even nuclei and weakening of  neutron  pairing.
This  feature  has  been  corroborated  by  Shell  Model (SM) and
Cranked Hartree Fock Bogoliubov (CHFB) results.

The  empirical  neutron  and  proton  pairgaps are related to the
odd-even staggering of binding energies \cite{Bertsch}. They  are
defined as

\vspace{-0.2cm}
\begin{eqnarray}
\Delta_n(Z,N) &=& {\frac{\pi_N} 2} [B(Z,N-1)-2*B(Z,N)+B(Z,N+1)]
\nonumber \\
%&&-2*B(Z,N)] \nonumber\\
 \Delta_p(Z,N) &= &{\frac{\pi_Z} 2} [B(Z-1,N)-2*B(Z,N)+B(Z+1,N)]
\nonumber
\end{eqnarray}
where  B(Z,N)  is  the  binding energy \cite{mass} of the nucleus
with Z protons and N neutrons. The factor depending on the number
parity ${\pi_N}$ (${\pi_Z}$) is chosen so that the  pairing  gaps
are  all  positive.  To exclude the effect of variation of single
particle energies in the staggering, the pairing gaps for  odd  N
(Z)'s  have  been  considered  as  the  neutron  (proton) pairing
energies. The neutron (proton) pairing energy at an even N (Z) is
estimated from  the  average  of  corresponding  values  for  two
neighbouring odd N (Z) nuclei.

\begin{figure}{~}
\vspace{2cm}
\includegraphics{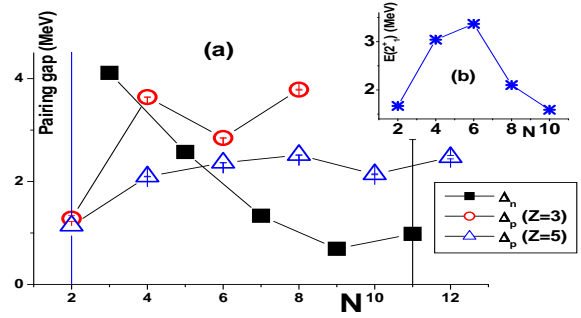}
\vspace{1.2cm}
\caption  {\label{figbe}  Variation  of  (a)  $\Delta_n$  for odd
isotopes of Be, $\Delta_p$ for odd isotopes  of  Li  and  B,  (b)
$E(2^+_1)$  energies  for  even  isotopes of $Be$ with increasing
neutron number (N).}

\end{figure}

In  Be  (Z=4)  isotopes,  the  N=8  shell  gap quenching has been
observed  \cite{be}.  Comparison  of  Figs.~\ref{figbe}a  and   b
reveals  how  $2^+_1$  states  are  built  in  Be  isotopes  as N
increases. In $^6Be$, this state is built by breaking a weak $\pi
1p_{3/2}$ proton pair. In $^{8,10}Be$, both protons and  neutrons
can  contribute.  The $2^+_1$ state in $^{10}Be$ can be generated
by breaking a neutron pair in $\nu 1p_{3/2}$ and exciting one  of
them  to $\pi 1p_{1/2}$. As a result the $2^+_1$ energy increases
(Fig.~\ref{figbe}b). With increasing neutron number, the  neutron
pairing decreases with increase in proton pairing. At around N=8,
the  proton pairing becomes stronger (Fig.~\ref{figbe}a) by about
2 MeV. In $^{12}Be$, instead of breaking  a  strong  proton  pair
($\Delta_p  \simeq$  3.1  MeV)  in $\pi 1p_{3/2}$ or exciting two
neutrons across the N=8 shell closure to $\nu 1d_{5/2}$, the most
favourable option to generate a 2$^+$ state,  is  to  deform  the
system  (nuclear  Jahn  Teller effect \cite{rein}) and bring down
the energy of the $\nu 1d_{5/2}$ orbit. The orbit below the  gap,
i.e  the  1/2[101]  Nilsson orbit originating primarily from $\nu
1p_{1/2}$ at low deformations,  is  insensitive  to  increase  in
deformation  compared  to  the downslope of the 1/2[220] orbit of
$\nu 1d_{5/2}$ \cite{nilsson}. This results in closing of the N=8
shell gap for  $^{12}Be$  isotope.  The  single  particle  energy
($spe$)  difference  $\nu  (1p_{1/2}-1d_{5/2})$  in  Be,  at N=8,
estimated from difference of $\Delta_n$ of relevant  even  N  and
(N+1)  \cite{Bertsch}  isotopes  is $\approx$ 1.8 MeV compared to
$\approx$ 3.9 MeV in corresponding C isotopes providing  evidence
of shell erosion. The weakening of the neutron pairing ($\Delta_n
\simeq$  0.8  MeV  at  N=10)  further  decreases  the  $E(2^+_1)$
energies of $^{14}Be$ without substantial increase in deformation
\cite{be}.

\begin{figure}{~}
\vspace{2cm}
\includegraphics{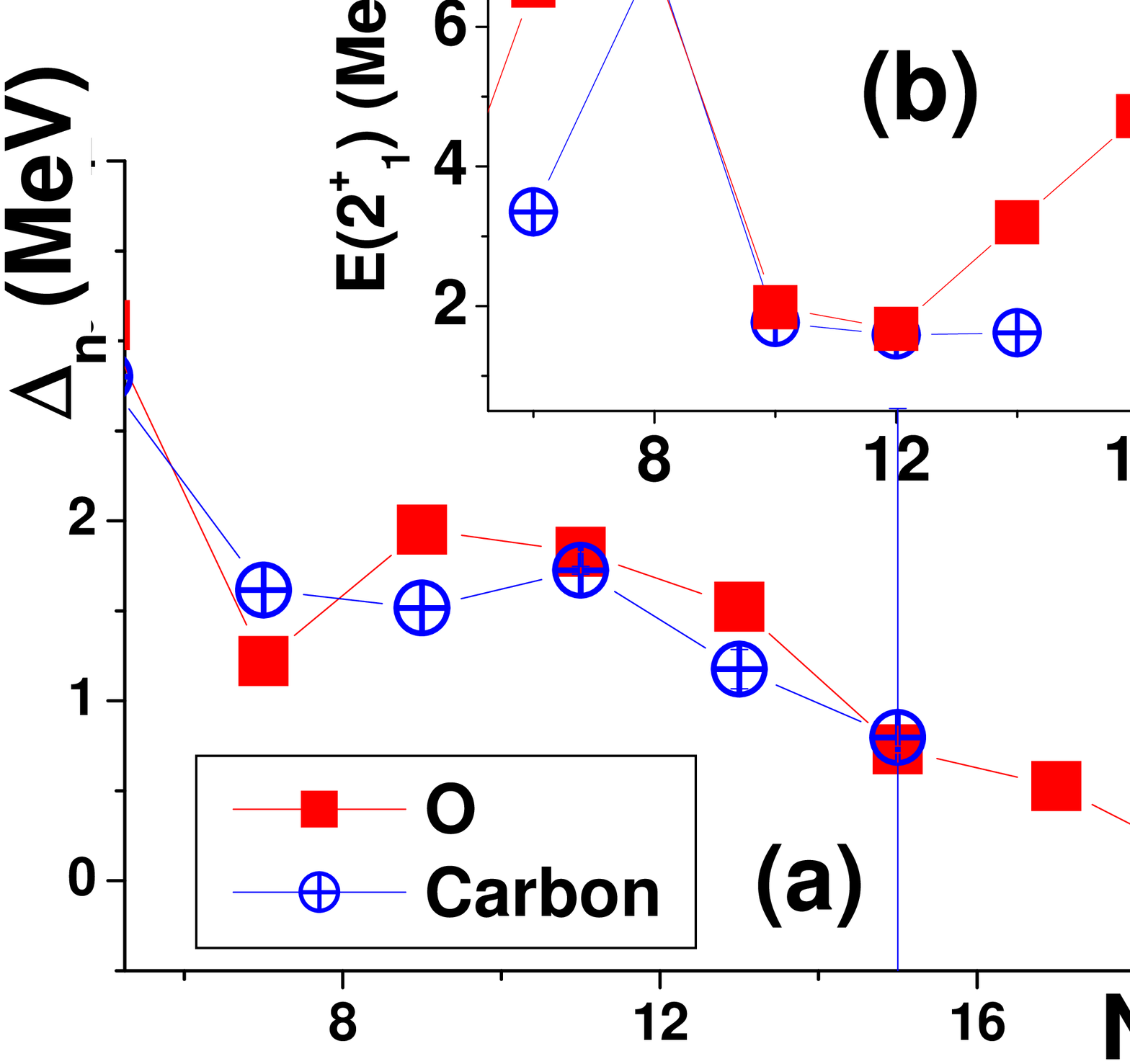}\vspace{1.2cm}
\caption  {\label{figco}  Variation  of  (a)  $\Delta_n$  for odd
isotopes, (b) $E(2^+_1)$ energies for even isotopes, of  $C$  and
$O$ with increasing neutron number (N).}

\end{figure}

For  C  and O isotopes at N=8, the proton pairing is nearly equal
to  the  neutron  pairing.  Fig.~\ref{figco}a  shows   that   the
$\Delta_n$'s for carbon and oxygen isotopes decrease with N above
N=11.  For  the  closure of the neutron $1d_{5/2}$ and $2s_{1/2}$
subshells at N=14 and 16, the semi-magic oxygen isotopes  exhibit
shell  closures at $^{22,24}O$, respectively (Fig.~\ref{figco}b).
The presence of two valence proton holes inhibits  the  occurrence
of  shell  closure  for the carbon isotopes. The $spe$ difference
$\nu (1d_{5/2}-2s_{1/2})$ and $\Delta_n$ at  N=14,  are  $\simeq$
(2.7,  1.1)  MeV  and  (1.3,  1.0)  MeV  for  O  and  C isotopes,
respectively. In $^{24}O$, $\nu (2s_{1/2}-1d_{3/2})$ is  $\simeq$
3.9  MeV  compared  to  $\Delta_n$ $\simeq$ 0.6 MeV. However, the
weakening of pairing in carbon  isotopes  is  manifested  through
decrease in its E($2^+_1$) values.

\begin{figure}{~}
\vspace{2.cm}
\includegraphics{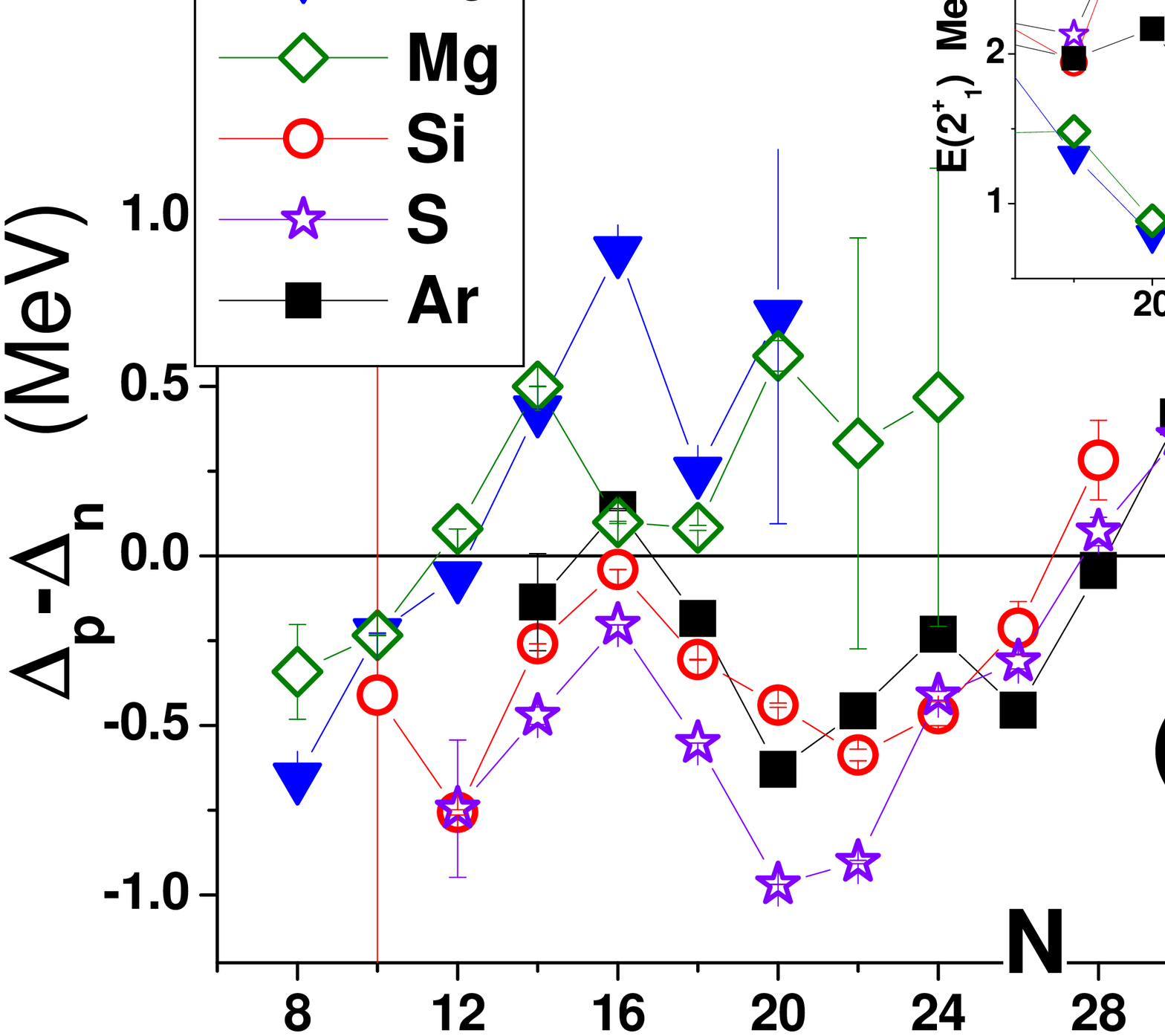}\vspace{1.cm}
\caption {\label{figsis} Variation of (a) $\Delta_p$ - $\Delta_n$
and  (b)  $E(2^+_1)$  energies  for  even isotopes of $Ne$, $Mg$,
$Si$, $S$ and $Ar$ with increasing neutron number (N).}

\end{figure}

Another  prominent example of the erosion of magicity is observed
at N = 20 for Z = 10, 12.  In  Fig.~\ref{figsis}a,  variation  of
$\Delta_p$  -  $\Delta_n$  (=$\Delta$),  the  difference  between
proton and neutron pairing energies with increasing  N  has  been
shown.  For  Ne  and Mg isotopes, across N=20, the proton pairing
shows an increase with corresponding decrease in neutron  pairing
($\Delta$  positive).  The situation is similar to Be isotopes at
N=8. The Nilsson orbitals involved are upsloping 3/2[202] from  $
\nu  1d_{3/2}$  and  relatively  faster downsloping 1/2[330] from
$\nu 1f_{7/2}$.  So  onset  of  deformation  along  with  reduced
neutron  pairing  depresses the $E(2^+_1)$ in N=20 isotopes of Ne
and Mg without substantial increase in B(E2)s. The  weakening  of
the  neutron pairing continues for higher N with further decrease
in the $E(2^+_1)$ energies of $^{32}Ne,~^{34}Mg$. For Si,  S  and
Ar  nuclei,  below  N=28,  $\Delta$ is negative (except for Ar at
N=16),  indicating  stronger  neutron  pairing.  However,  in  Si
nuclei,  $\Delta$ crosses the zero line and becomes positive from
N$\simeq$ 27, while in S and Ar crossing is at N=28. So Si  shows
an   erosion   of  the  N=28  shell  closure  manifested  through
depression in E($2^+_1$) energy (Fig.~\ref{figsis}b) due to onset
of deformation where neutrons from upsloping 7/2[303]  from  $\nu
1f_{7/2}$   can  be  excited  to  downsloping  1/2[321]  of  $\nu
2p_{3/2}$, with 5/2[312] from $\nu 1f_{7/2}$ being insensitive to
increase in deformation \cite{nilsson}. In  S  and  Ar  isotopes,
erosion   is   not  distinct,  but  beyond  N=28,  depression  in
E($2^+_1$) with slower increase in B(E2) is observed \cite{Ar}.

\begin{figure}{~}\vspace{2.2cm}
\includegraphics{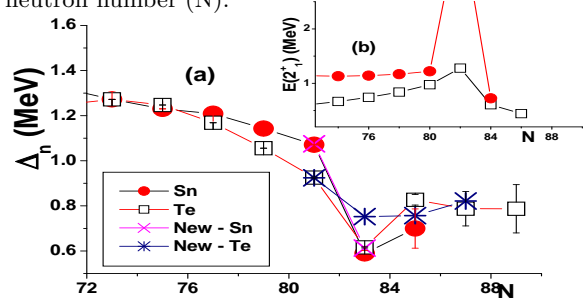}
\vspace{.7cm}
\caption  {\label{figsnte}  Same as Fig.~\ref{figco} for $Sn$ and
$Te$. "New" indicates the $\Delta_n$ calculated from most  recent
mass measurements \cite{hakala}. }

\end{figure}

The  Sn region is even more interesting, showing how new features
develop as one moves far away from stabilty for heavy nuclei. For
neutron-rich isotopes of Sn and  other  heavier  nuclei,  neutron
pairing shows reduction. However, the proton pairing is generally
smaller  than  or  close to neutron pairing for these nuclei. The
region above neutron-rich  doubly  closed  $^{132}Sn$  has  great
similarities \cite{blom} with that above doubly closed $^{208}Pb$
on  the  stability.  The  even $Sn$ isotopes have nearly constant
$E(2^+_1)$ ($\simeq$ 1200 keV)  values  for  A=102-130.  This  is
attributed  to  the near degeneracy of the neutron $1g_{7/2}$ and
$2d_{5/2}$  single  particle  orbitals  which  enhances   pairing
correlations.  Similarly,  there  is a general belief that for Sn
isotopes beyond $^{132}Sn$, there will be enhancement of  pairing
resulting  in  a  constant  $0^+_1-2^+_1$ spacing, as revealed in
calculations  with  realistic  interactions   \cite{sah:2,karta}.
However,   suddenly  at  N=84,  for  $^{134}Sn$,  the  $E(2^+_1)$
decreases   to   726   keV.   This   decrease    of    $E(2^+_1)$
(Fig.~\ref{figsnte})  is  clearly  correlated  to the decrease in
pairing manifested through the reduction in $\Delta_n$  ($\simeq$
1.2  MeV  at  N=81  to  0.6  MeV  at  N=83).  This observation is
strengthened  by  the  most  recent  precise  mass   measurements
\cite{hakala}.  On  the  other hand, for Pb isotopes above N=126,
the $\Delta_n$ values are  very  similar  to  those  below  N=126
($\simeq$ 0.68,0.32,0.62 MeV at N=123,125, 127, respectively) and
so  a  near  constancy  of  the $E(2^+_1)$s below and above N=126
closure  (0.899,  0.803,  0.800  MeV   for   N=122,   124,   128,
respectively)  has  been observed. Pairing for different isotopes
of Te, Xe and Ba nuclei has been estimated. At  N=82,  $E(2^+_1)$
(1.279,  1.313  and 1.435 MeV, respectively) for Te, Xe and Ba is
$\simeq 2 \Delta_p$ ($\Delta_p$ $\simeq$ 0.66, 0.82 and 0.87 MeV,
respectively). Compared to that in N=78 ($\Delta_n$ $\simeq$  1.1
MeV for Te, Xe and Ba), beyond N=82, reduction in neutron pairing
($\Delta_n$  $\simeq$ 0.72, 0.83, 0.84 MeV at N=84 for Te, Xe and
Ba,  respectively)  is  observed.  It  gives  rise  to  depressed
$E(2^+_1)$   (0.606,  0.589  and  0.602  MeV,  respectively)  and
relatively hindered B(E2)s depending on the degree of  reduction.
Weaker $\Delta_p$ at N=82 favour easy generation of $2^+_1$ state
without any need for neutron shell erosion.

Analysing   the   empirical   observations  of  the  features  of
neutron-rich nuclei one can frame the following rules.

\vspace{-.3 cm}
\begin{itemize}
\item{}  For  proton - magic nuclei, reduction in neutron pairing
will give rise to

\vspace{-.3 cm}
\begin{itemize}
\item{}
anomalous   decrease   in  E($2^+_1$)  with  slower  increase  in
deformation for isotopes with neutrons in unfilled subshells. For
normal pairing observed near  stability,  similar  value  of  low
E($2^+_1$)'s  will  indicate  higher  value  of  B(E2)  or larger
deformation.

\vspace{-.2 cm}
\item{}
appearance  of  new shell closures manifested through enhancement
of E($2^+_1$) for isotopes with filled up sub-shells. The  energy
gap  between  the sub - shells must be substantially greater than
the neutron pairing energy.

\end{itemize}
\vspace{-.3  cm}
\item{}  For  nuclei  with  a  few  valence  particles near shell
closure having unfilled proton sub-shell, reduced neutron pairing
will lead to

\begin{itemize}
\vspace{-.3 cm}
\item{}  anomalous decrease in E($2^+_1$) with slower increase in
deformation for isotopes with neutrons in unfilled sub-shells,

\vspace{-.2 cm}
\item{}  onset  of deformation for N$_{magic}$ nucleus, resulting
in  erosion  of  shell  gap  if  proton  pairing   shows   strong
enhancement.  This  will  be  usually  observed  for lighter mass
nuclei.

\end{itemize}

\end{itemize}

To  have  a  quantitative  estimation of the effect of pairing on
depression  in  the  $E(2^+_1)$  values  and  on  the  onset   of
deformation,  shell  model  calculation  has been done for oxygen
isotopes. It has been shown that with reduction of $0^+$ diagonal
two  body  matrix  elements  of  Bonn  A  realistic   interaction
\cite{physrep},  $E(2^+_1)$  energies  of  $^{18-22}O$  decreases
without any increase in the  B(E2)  values  with  improvement  in
prediction  of  a  shell  gap  at  N=14, usually not predicted by
Bonn-A. The same is also true for even-even  Sn  isotopes  beyond
$^{132}Sn$ as also discussed in Ref. \cite{sah:2}.

\begin{figure}{~}
\vspace{3.cm}
\includegraphics{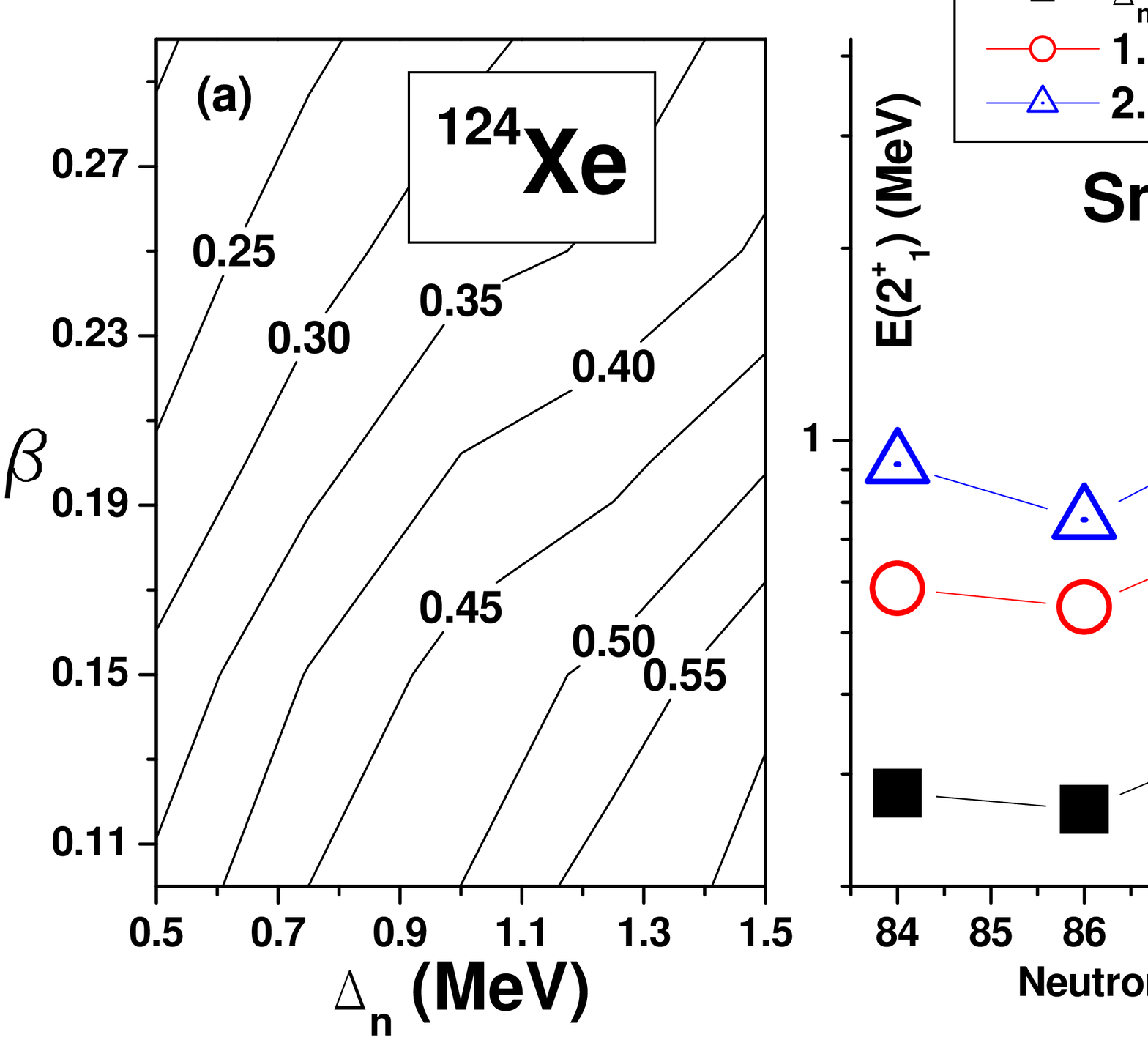}
\vspace{0.5cm}
\caption {\label{figxesn}(a) The equi- E$(2^+_1)$ energy contours
are  shown  as  a  function  of deformation ($\beta$) and neutron
pairing  gap  ($\Delta_n)$.  (b)  The  variations  in  E$(2^+_1)$
energies  for  $^{134-140}Sn$  with  reduction in neutron pairing
gap.}

\end{figure}

From  cranked  Hartree-Fock-Bogoliubov (CHFB) calculations with a
pairing plus quadrupole  Hamiltonian,  equi-energy  contours  for
$E(2^+_1)$  energies  in  $^{124}Xe$,  well  studied  within this
formalism \cite{prc} have been  plotted  in  Fig.~\ref{figxesn}a,
for variations of neutron pairing gap and quadrupole deformation.
For  all  these  contours  the  proton  pairing gap has been kept
constant at 1.5 MeV, which  is  obtained  from  the  experimental
odd-even mass difference. These contours are useful to understand
the  effect of variation of pairing on E$(2^+_1)$ for fixed value
of  deformation  and  vice-versa.  For  example,   the   measured
E$(2^+_1)$ value ($\simeq$ 350 keV) for $^{124}$Xe can correspond
to   two   extreme   combinations  of  neutron  pairing  gap  and
deformation.  They  are  ($\Delta_n$=1.4  and  $\beta$=0.30)  and
($\Delta_n$=0.5   and   $\beta$=0.11)  (Fig.~\ref{figxesn}a).  It
implies that same depressed value of energy can result either due
to increase in  deformation  with  usual  pairing  or  relatively
smaller deformation with reduced of pairing.

Self-consistent  solutions  have been obtained to investigate how
deformation changes with decreasing pairing and vice - versa  for
$^{124}Xe$.  The  results  are shown in Table I. It is found that
even without  any  change  of  quadrupole  interaction  strength,
reduction  in  neutron  pairing  strength favours a slow onset of
deformation (Part (a) of Table I). It is also evident  that  same
depressed  E($2^+_1)$  energy ({\it e.g.}, $\simeq $ 120-150 keV)
can result either from a reduced pairing  ($\Delta_n  \simeq$  0)
and relatively smaller deformation ($\beta_2 \simeq 0.26$) (Table
Ia),  or  from  enhanced  deformation ($\beta_2 \simeq 0.37$) for
stronger pairing ($\Delta_n \simeq$ 0.6 MeV) (Table Ib).

\begin{table}
\caption{Effects  of  (a)  decrease  in  neutron pairing strength
(g$_n$) and (b)  increase  in  quadrupole  interaction  strength,
$x_2$,  on  ground state deformation, pairing gaps and E(2$^+_1$)
values of $^{124}Xe$. In  part  (a),  the  $x_2$  value  is  kept
constant  at 72.0, in (b), the value of g$_n$ is 20.0. The proton
pairing strength (g$_p$) is kept constant at 26.0  in  all  these
calculations. The energies are in MeV.}

\begin{ruledtabular}
\begin{tabular} {cccccc}
& $g_n$  &
$\beta_2$ & $\Delta_p$ & $\Delta_n$& E(2$^+_1$)  \\
 \hline
$(a)$&25&	0.087&	1.368&	1.875&	0.773\\
&20&	0.182&	1.222&	1.047&	0.304\\
&15&	0.230&  1.171&	0.433&	0.172\\
&10&	0.261&	1.168&	0.00003&	0.125\\
&5&	0.261&	1.168&	0.00001&	0.114\\
 \hline
&$x_2$  &
$\beta_2$ & $\Delta_p$ & $\Delta_n$& E(2$^+_1$)  \\
 \hline
$(b)$&72&0.182&	1.222&	1.047&	0.304\\
&77&0.256&1.167 &0.834&0.233\\
&82&0.366&1.116&0.605&0.151\\
\end{tabular}
\end{ruledtabular}
\end{table}
To  test  the  observation  for proton-magic nuclei, we have done
another set of calculations for $Sn$ isotopes  above  $^{132}Sn$.
Single  particle  energies  have  been  obtained from shell model
Hamiltonian in Refs. \cite{sah:1,sah:2,epja}. The single particle
energies for the $\pi 1g_{9/2}$ and $\nu 1h_{11/2}$ orbitals have
been chosen with sufficient care  so  that  $Sn$  isotopes  above
A=132  do  not show any protons or neutron excitations across the
Z=50 and N=82 shell closures, respectively. The remaining  single
particle  energies  are  calculated  using  Nilsson  prescription
\cite{nil} with proper normalisation.

From  the  Fig.~\ref{figxesn}b,  it  is  observed that as pairing
energy  for  neutron  decreases,  the  shell  closure   at   N=90
\cite{sah:1}  manifested through sudden jump in E($2^+_1$) energy
of  $^{140}Sn$  compared  to  that  in  $^{138}Sn$  becomes  more
pronounced.  For  stronger  pairing  the  E($2^+_1$)  energies of
semi-magic nuclei show  much  slower  variation  with  increasing
neutron  numbers,  as  also  observed  for  isotopes  of Sn below
$^{132}Sn$.

From  the  experimental data on odd-even staggering of masses and
SM and CHFB calculations, we have shown  that  pairing  plays  an
important  role in the shell evolution of neutron-rich nuclei. We
have identified a few distinctive features of neutron-rich nuclei
based on our analysis of the  empirical  data.  For  proton-magic
nuclei,  due to reduction of neutron pairing away from stability,
each sub-shell closure will be manifested as a new shell closure,
if energy gap between sub-shells is greater than $\Delta_n$.  For
other  neutron-rich  isotopes  for  which  proton numbers are not
magic, the E($2^+_1$) energies at  each  existing  neutron  shell
closure  will  show  dramatic decrease depending on the extent of
reduction  in  neutron  pairing.  This  reduction  will  also  be
manifested  through  slow onset of deformation at shell closures.
But these values of the deformation will not  correspond  to  the
same  variation  as  seen  on  the  stability.  The  increase  in
deformation with decrease in energy of the $2^+_1$ state will  be
much  slower  for  reduced  pairing.  For  lighter  mass  nuclei,
increase in proton pairing has been found to play  a  significant
role   in   the  observed  shell  erosion.  This  work  therefore
conclusively shows the role of variation in pairing as a function
N in nuclear shell evolution  of  neutron  -  rich  nuclei.  This
method  of  analysis  is  also  applicable  to exotic proton-rich
nuclei. The microscopic origin of variation  in  pairing  with  N
constitutes  a  separate  important issue. Our work also warrants
more precise  measurement  of  masses  away  from  stability  and
analysis of mass data to search for newer isotopes for which such
shell evolution can be expected.

\end{document}